\documentclass[trackchanges,twocolumn]{aastex701}
\newcommand{\eps}[1]{\log\varepsilon_{\rm #1}}
\newcommand{\kms}{km\,s$^{-1}$}
\newcommand{\iso}[2]{\mbox{$^{#1}{\rm #2}$}}

\newcommand{\Teff}{T_{\rm eff}}
\newcommand{\logg}{log\,$g$}
\newcommand{\Eexc}{E_{\rm exc}}

\newcommand{\eu}[5]{\mbox{$#1\,^#2{\rm #3}^{#4}_{\rm #5}$}}

\begin{document}

\title{Understanding the origin of silver in metal-poor stars based on \\ the non-local thermodynamic equilibrium analysis of \ion{Ag}{1} lines\footnote{Based on observations made with the telescopes of European Southern Observatory (ESO)}}

\author[orcid=0000-0002-3188-8263,sname='Mashonkina']{Lyudmila Mashonkina}
\affiliation{Institute of Astronomy, Russian Academy of Sciences}
\email[show]{lima@inasan.ru}  

\author[orcid=0000-0002-4174-1505,sname='Sitnova']{Tatyana Sitnova}
\affiliation{Institute of Astronomy, Russian Academy of Sciences}
\email{sitamih@gmail.com}  

\author[orcid=0000-0002-8889-7283,sname='Yakovleva']{Svetlana A. Yakovleva}
\affiliation{Institute of Physics, Herzen University}
\email{cvetaja@gmail.com}  

\author[sname=Vasil'eva]{Vera A. Vasil'eva}
\affiliation{Institute of Physics, Herzen University}
\email{veravasilieva2000@gmail.com}

\begin{abstract}

Silver is a poorly observed neutron-capture element, and understanding its Galactic production history requires accurate abundances for stellar samples covering a broad metallicity range. We refined the method for calculating stellar Ag abundances by considering the non-local thermodynamic equilibrium (NLTE) line formation for \ion{Ag}{1}. A comprehensive model atom of silver was built from the atomic data available in the literature, including recent data on inelastic \ion{Ag}{1} + \ion{H}{1} collisions. We derived abundances using the high-resolution spectra and the synthetic spectrum method with the one-dimensional model atmospheres. From two \ion{Ag}{1} lines in the solar spectrum we obtained the NLTE abundances $\eps{\odot,Ag}$ = 1.15 and 1.10 for different treatments of collisions with hydrogen atoms. The NLTE abundances of Ag were determined for the first time for 41 stars in the $-2.74 \le$ [Fe/H] $\le -0.56$ metallicity range. The NLTE abundance corrections for \ion{Ag}{1} 3382~\AA\ were found to be larger for the giants (up to 0.92~dex) than the dwarfs (up to 0.51~dex). In NLTE, Ag in our sample stars is enhanced relatively to Fe and the [Ag/Fe] Galactic trend is very similar to that for [Eu/Fe]; the [Ag/Eu], [Ag/Ba], and [Ag/Sr] abundance ratios of the r-II stars CS22892-052 and CS31082-001 agree very well with the r-process model of Prantzos~et~al. (2020); the $-1.6 \lesssim$ [Fe/H] $\le -0.56$ stars have close-to-solar [Ag/Eu] ratios. We conclude that production of silver was dominated by the main r-process throughout the metallicity range we investigated, with the main s-process contribution of not larger than 20\%\ for [Fe/H] $\gtrsim -1.6$.

\end{abstract}


\section{Introduction}

Since the classical paper by \citet{BBFH1957}, one commonly believes that chemical elements beyond the Fe group are produced in neutron capture (n-capture) nuclear reactions, from the slow (s-) process and the rapid (r-) process in differing proportions for different isotopes.
The mechanisms and sites of the s-process are relatively well understood \citep{Kappeler2011}. It operates at neutron number densities of $N_n \sim 10^6 - 10^7$~cm$^{-3}$ and is represented by the main and weak s-process components. 
The main s-process occurs in low- and intermediate-mass ($0.9-8 M_\odot$) stars on their  asymptotic giant branch (AGB) phase and is able to produce nuclei from gallium (Ga, $Z$ = 31) up to bismuth (Bi, $Z$ = 83).
The other s-process component is associated with the He-burning cores of massive ($\sim 20 M_\odot$) stars and produces nuclei with an atomic mass of up to $A \sim$ 90 more effectively than the heavier ones \citep{2012A&A...538L...2F}.
The r-process takes place at extremely high $N_n > 10^{24}$~cm$^{-3}$, which can be reached in explosive conditions occurring at final evolutionary stages of massive ($> 8 M_\odot$) stars, such as neutron star mergers (NSM) and magnetorotational supernovae \citep[see][for a review]{2021RvMP...93a5002C}. Despite support of the NSM scenario for the gravitational-wave event GW170817 from the detection of its electromagnetic signal as the $\gamma$-ray burst GRB 170817A and kilonova AT2017gfo \citep{2017ApJ...848L..12A}, 
the identification of a fully convincing astrophysical site is still under discussion
 \citep{2023ApJ...943L..12K}. The n-capture chain of the r-process can proceed beyond Bi.


The relative contribution of the r- and s-process to production of a given heavy element varied during the Galaxy history because the lifetimes of the r- and s-nucleus sources are different. This paper focuses on silver. For the Solar System matter, 
the s-process calculations of \citet{Prantzos2020} predict that 20.7\%\ of the silver originate from the s-process,
 and thus the remaining 79.3\%\ from the r-process. 
For the earlier Galactic epochs, the r/s ratio can be 
 estimated from observations of silver in stellar samples covering a broad
metallicity range.

Silver is a poorly observed element in late type stars. 
With a nuclear charge of $Z$ = 47 and a number of neutrons ($N_n$ = 60 and 62 for the $\iso{107}{Ag}$ and $\iso{109}{Ag}$ isotopes) that is far from the magic numbers $N_n$ = 50, 82, and 126, Ag has a lower abundance than Sr and Ba, for example. Furthermore, Ag is represented in stellar spectra by two lines of \ion{Ag}{1} that are located at $\lambda <$ 3400~\AA, in the spectral range, which is rather crowded even in metal-poor (MP, [Fe/H]\footnote{In the classical notation, where [X/Y] = $\log(N_{\rm X}/N_{\rm Y})_{star} - \log(N_{\rm X}/N_{\rm Y})_{\odot}$ for each pair of elements X and Y.} $< -1$) stars.
Silver abundances were measured for strongly r-process enhanced ([Eu/Fe] $> 1$, [Ba/Eu] $< 0$) stars \citep{Sneden2003,hill2002,2017A&A...607A..91H}, which are referred to as r-II stars, according to the definition of \citet{HERESI}. The first abundance determinations for Ag in a large sample of MP stars were reported by \citet{2012A&A...545A..31H}, who concluded that the Galactic trends of heavy element ratios involving Ag suggest a contribution to the Ag production from a second or weak r-process. The weak r-process \citep[see, for example,][]{2017JPhG...44e4003B} is associated with the $\alpha$-rich freeze-out \citep{1992ApJ...395..202W} that was introduced for explanations of the high abundance ratios between the lighter heavy elements, such as Sr, Y, Zr, and the heavier elements, such as Ba and Eu, 
 observed in some very MP (VMP, [Fe/H] $< -2$) stars. 
 The statistics of stellar Ag measurements was increased by \citet{2015A&A...579A...8W,2024A&A...690A.262B}, and 
  \citet{2025ApJ...991..112H}.

 The silver abundance determinations made so far were all based on the assumption of local thermodynamic equilibrium (LTE). \citet{2012A&A...545A..31H} found an abundance difference between dwarfs and giants of similar metallicity and suggested that the departures from LTE might be a possible explanation. Indeed, with the ionization energy $\chi$ = 7.576~eV, silver is ionized in physical conditions that are typical for the atmospheres of MP stars and can be subject to overionization, similarly to other neutrals, for example, \ion{Fe}{1} and \ion{Pd}{1}. Calculations based on the non-local thermodynamic equilibrium (NLTE) line formation for \ion{Ag}{1} were first performed by \citet{2026A&A...711A.155C} for the Sun. They found positive abundance differences between NLTE and LTE, at the level of 0.2~dex in the solar plane-parallel (1D) and 0.27~dex in the three-dimensional (3D) model atmosphere. 
 For any other star, the NLTE calculations for \ion{Ag}{1} are still absent.
  
 This paper aims to restore the evolution of the Galactic silver abundance and to uncover  contributions of the r- and s-process to Ag in different Galactic epochs based on accurate stellar NLTE abundances and the r/s-process predictions of \cite{Prantzos2020}. We set the following tasks.
 \begin{itemize}
 	\item Construction of a comprehensive model atom of \ion{Ag}{1} from the radiative and collisional data available in the literature and the rate coefficients for the excitation and ion pair production in the \ion{Ag}{1} + \ion{H}{1} collisions obtained from our own calculations.
 	\item Determinations of the silver NLTE abundances for the Sun and the sample of MP stars, for which the NLTE abundances of Pd were earlier derived by \citet{pd_nlte}.
 	\item Analysis of the Galactic trends of abundance ratios among Ag and the elements, which are tracers of different n-capture processes, that is the r-, main s-, and weak s-process.
 \end{itemize}
 
The paper is organized as follows. Section~\ref{Sect:NLTE} introduces the model atom for silver. In Sect.~\ref{Sect:sun}, it is applied to determining the NLTE abundances from the solar \ion{Ag}{1} lines. Section~\ref{Sect:abund} describes our stellar sample, the sources of observed spectra and atmospheric parameters, and determinations of stellar Ag abundances. Galactic trends of the elemental ratios are discussed in Sect.~\ref{Sect:trends}. 
Our conclusions are given in Sect.~\ref{Sect:conclusion}. 


\section{Model atom for silver} \label{Sect:NLTE}

\subsection{Energy levels}

\begin{figure}   
\plotone{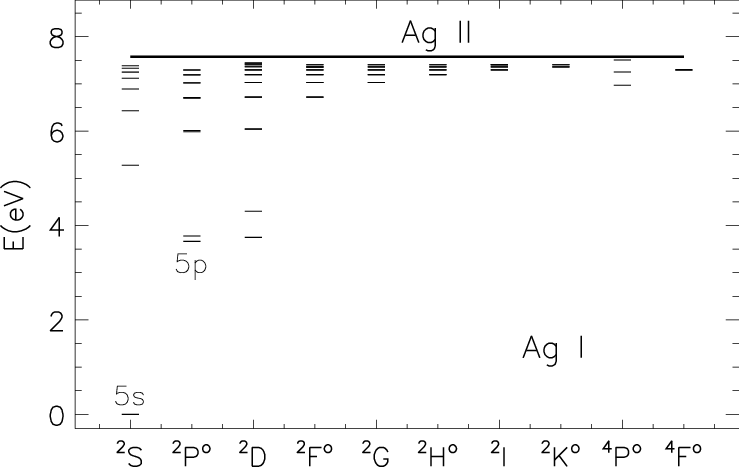}
\caption{Energy levels of \ion{Ag}{1}, which were used in building the model atom.
\label{fig:atom}}
\end{figure}

The atomic model was built using 60 levels of \ion{Ag}{1} from the database of the National Institute of Standards and Technology (NIST)\footnote{https://physics.nist.gov/asd} \citep{NIST19} and 36 levels predicted in calculations of the \ion{Ag}{1} atomic structure \citep[][files c4700e.log and c4700o.log from April 11 of 2011]{Kurucz2026}, which were not detected in laboratory experiments so far. All these predicted levels lie above the excitation energy $\Eexc$ = 7.2~eV. The high-excitation levels provide a close collisional coupling to the first-ionization stage, \ion{Ag}{2}, and thus play an important role in the statistical equilibrium of \ion{Ag}{1}. The energy levels, which were used in constructing the model atom, are displayed in Fig.~\ref{fig:atom}.

 Fine splitting levels of the two lowest excited terms, \eu{4d^{10} 5p}{2}{P}{\circ}{} and \eu{4d^9 5s^2}{2}{D}{}{}, were explicitly included in the model atom. For $\Eexc >$ 7.2~eV, the levels with a low energy separation ($< 0.01$~eV) and of the same parity were combined into superlevels. 
For each superlevel, we calculated its energy from energies of individual levels and took their statistical weights into account. 

Singly-ionized silver is represented in the model atom by the ground state.
The other levels of \ion{Ag}{2} lie all above $\Eexc$ = 4.85~eV and, for the temperatures relevant to this study, contribute only a little to the partition function, for example, 0.1\%\ for $T$ = 6000~K.

\subsection{Radiative transitions}

The final model atom includes 98 allowed bound-bound (b-b) transitions. Their oscillator strengths were taken from the calculations of \citet[][file gf4700.pos from April 11 of 2011, hereafter, K11]{Kurucz2026}. 
We note that log~$gf$(K11) = $-0.299$ and 0.013 for the resonance transitions \eu{5s}{2}{S}{}{} -- \eu{5p}{2}{P}{\circ}{1/2,3/2} agree within 0.03~dex with the experimental values from \citet{1990ZPhyD..16...87C}. 
We compared $gf$(K11) with recent calculations of \citet[][JS26]{2026A&A...709A..31J}, who applied the multiconfiguration Dirac-Hartree-Fock and relativistic coupled-cluster methods. 
For 57 transitions, in common, the mean difference between JS26 and K11 is 0.06$\pm$1.04~dex. The large dispersion is mostly due to the four transitions with $gf$(JS26) -- $gf$(K11) of 2.1 to 4.2~dex, by absolute value.
In Sects.~\ref{Sect:sun} and \ref{Sect:abund}, we inspect an influence of replacing $gf$(K11) with $gf$(JS26), where available, on the NLTE results.
 
The photoionization cross-sections were calculated in the hydrogenic approximation using the effective principal quantum number instead of the principal quantum number. The effect of variations in the photoionization data on the NLTE results is discussed in Sect.~\ref{Sect:abund}.

\subsection{Collisional transitions}\label{Sect:coll}

The electron-impact excitation was computed with the semi-empirical formula of \citet{Reg1962} for the allowed transitions and with the effective collision strength $\Upsilon$ = 1 for the forbidden ones, because no accurate data are available. The formula by \citet{1962amp..conf..375S} was used to calculate the electron-impact ionization rates.

Inelastic processes in collisions with \ion{H}{1} atoms were treated by applying the rate coefficients from two different approaches. For excitation in the transitions between the lowest levels of \ion{Ag}{1} up to \eu{7s}{2}{S}{}{} and charge transfer associated with the same nine levels, \citet[][YV26]{ag_hyd} modified asymptotic model within the Born-Oppenheimer approach proposed by \citet{2013PhRvA..88e2704B}. These calculations use ab initio data for six low-excitation states in order to account for the fine structure of the two lowest excited terms \eu{4d^{10} 5p}{2}{P}{\circ}{} and \eu{4d^9 5s^2}{2}{D}{}{}. For including non-adiabatic regions at small internuclear distances into nuclear dynamics treatment, the probability currents method is employed.
The NLTE calculations performed with this data set are called variant YV26.

\begin{figure}   
\plotone{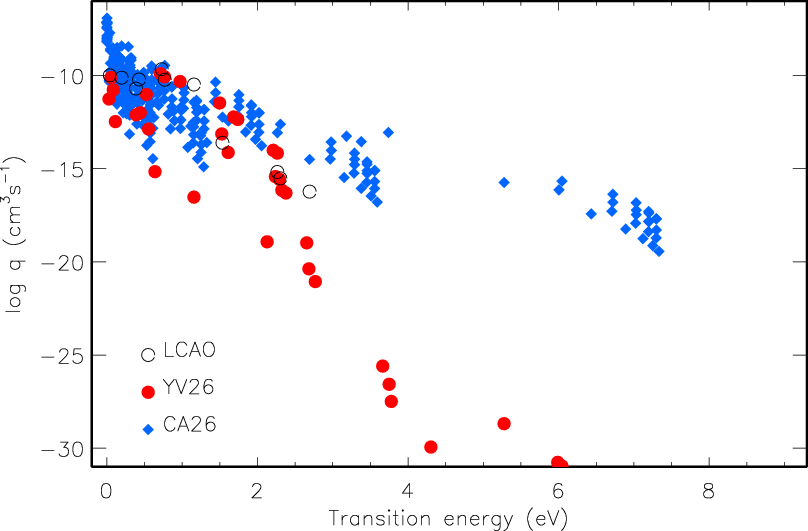}
\caption{Excitation rate coefficients, log~q, for the \ion{Ag}{1} + \ion{H}{1} collisions as a function of the transition energy from calculations of \citet[][filled circles, YV26]{ag_hyd}, 
\citet[][LCAO, open circles]{2026A&A...711A.155C}, and \citet[][LCAO + free electron model, diamonds]{2026A&A...711A.155C}. The data correspond to $T = 5000$~K. \label{fig:coll}
}
\end{figure}

The second data set was taken from \citet{2026A&A...711A.155C}. For 11 b-b and seven bound-free (b-f) transitions, they applied the LCAO (linear combination of atomic orbitals) approach by \citet{2016PhRvA..93d2705B}.
We found that  
the LCAO and YV26 rate coefficients are fairly consistent for both excitations (see Fig.~\ref{fig:coll}) and charge transfer.
 For excitation via neutral hydrogen collisions, \citet{2026A&A...711A.155C} also used the free electron model by \citet{2018A&A...616A..89A} and provided the rate coefficients for 178 transitions, in total. Our NLTE calculations based on the LCAO + free electron model rates from \citet{2026A&A...711A.155C} will be referred below to as variant CA26. 
  As seen in Fig.~\ref{fig:coll}, the YV26 and CA26 rates lie at the same level for $\Eexc <$ 2~eV, although the number of transitions in the CA26 calculations is much larger.

\section{Solar abundance of Ag} \label{Sect:sun}


\begin{deluxetable*}{ccccccc}
\tablewidth{0pt}
\tablecaption{Solar LTE and NLTE (variants CA26 and YV26) abundances, $\eps{}$ \label{tab:sun}}
\tablehead{
\colhead{ Line} & \colhead{$\Eexc$} & \colhead{log $gf$} & \colhead{Transition} &  \colhead{LTE} & \colhead{CA26} & \colhead{YV26} 
}
\startdata
 \ \ion{Ag}{1} 3280.68 \AA & 0.00 & --0.022 & \eu{5s}{2}{S}{}{1/2} -- \eu{5p}{2}{P}{\circ}{3/2} & 0.83 & 1.08 & 1.14 \\
 \ \ion{Ag}{1} 3382.89 \AA & 0.00 & --0.334 & \eu{5s}{2}{S}{}{1/2} -- \eu{5p}{2}{P}{\circ}{1/2} & 0.89 & 1.10 & 1.15 \\
\enddata
\end{deluxetable*}

We used the canonical solar parameters: the effective temperature $\Teff$ = 5780~K, the surface gravity \logg\ = 4.44, microturbulent velocity of $\xi_{t}$ = 0.9~\kms\ and the classical atmospheric model from the MARCS \citep{Gustafssonetal:2008} database\footnote{http://marcs.astro.uu.se}. 
Abundances were determined from the \ion{Ag}{1} 3280 and 3382~\AA\ lines in the intensity spectrum of the solar disk center by \citet{Delbouille1973}, 
using the synthetic spectrum method, that is, by fitting the theoretical to the observed spectrum. 

Everywhere in this study, the coupled statistical equilibrium (SE) and radiative transfer equations in a given atmospheric structure were solved with the {\sc detail} code \citep{Giddings81,Butler84}, where the opacity package was revised as described by \citet{mash_fe}. 
The synthetic intensity spectra were calculated with the code \textsc{siu} \citep{Reetz}. \textsc{siu} calculated the \ion{Ag}{1} lines by applying the departure coefficients from {\sc detail} and thus taking the NLTE effects into account, while other lines by assuming LTE.
The line list and the line atomic parameters were taken from the Vienna Atomic Line Database \citep[VALD,][]{2015PhyS...90e4005R}. In the synthetic spectra calculations, solar abundances recommended by \citet{Lodders2021} were adopted as an initial guess.

For the \ion{Ag}{1} lines, we adopted experimental $gf$-values from \citet{1990ZPhyD..16...87C}, see  Table~\ref{tab:sun}.
Silver is represented in nature by the isotopes $^{107}$Ag and $^{109}$Ag, and each \ion{Ag}{1} line is composed of the hyperfine splitting (HFS) components of both isotopes. Their wavelengths and relative intensities were taken from \citet{2012A&A...545A..31H}. When computing $gf$-values, we adopted the Solar System isotopic ratio $^{107}$Ag : $^{109}$Ag = 51.84 : 48.16 \citep{Lodders2021}. 

\begin{figure*}   
	\plottwo{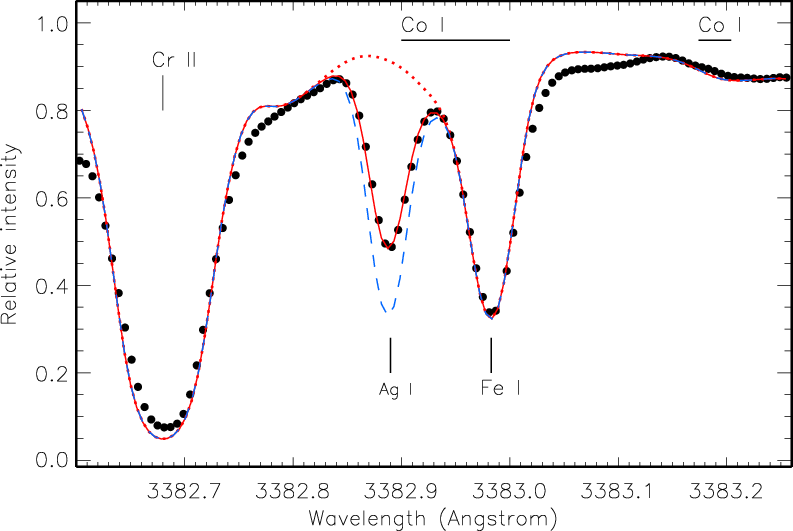}{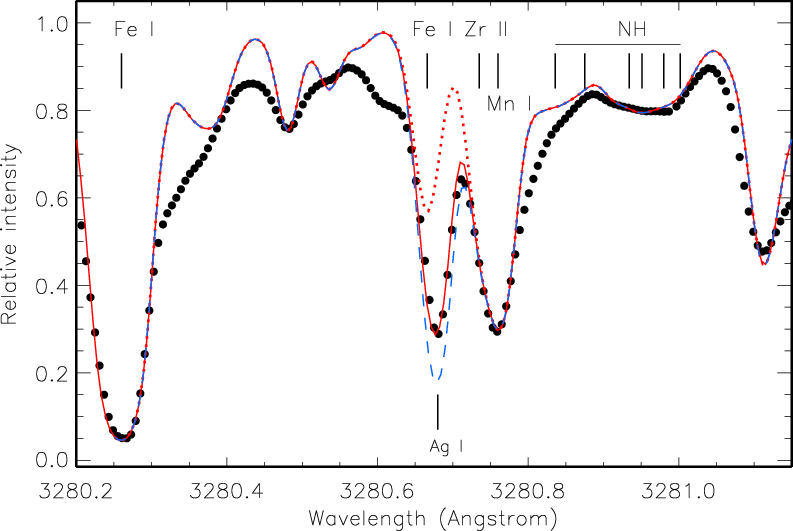}
	\caption{Best NLTE(YV26) fits (continuous curves) of the \ion{Ag}{1} lines in the intensity spectrum of the solar disk center \citep[][circles]{Delbouille1973}. The NLTE abundances we obtained are indicated in Table~\ref{tab:sun}. In each panel, the dashed curve is the LTE profile computed with the corresponding NLTE abundance, and the dotted curve is the synthetic spectrum without the contribution of the \ion{Ag}{1} line.
		\label{fig:sun}}
\end{figure*}

Of two lines, \ion{Ag}{1} 3382.89~\AA\ (Fig.~\ref{fig:sun}) is the more reliable abundance indicator. It is located in the overlapping wings of the strong lines \ion{Cr}{2} 3382.86~\AA\ and \ion{Ti}{2} 3383.76~\AA. In our computations, we included the \ion{Fe}{1} 3382.986~\AA\ line ($\Eexc$ = 3.017~eV, log~$gf = -2.25$) found by \citet{1966sst..book.....M} in the solar spectrum, but absent in the VALD database. The 3382.9~\AA\ blend is also contributed from the HFS components of the \ion{Co}{1} line, however, they affect the derived Ag abundance very little. With fixed abundances of Ti, Cr, Fe, and Co, the 3382.6--3383.3~\AA\ spectral range was fitted by treating the Ag abundance and the macroturbulent velocity as free parameters.

There exist problems with using \ion{Ag}{1} 3280.68~\AA\ in abundance determinations. As seen in Fig.~\ref{fig:sun}, this line overlaps with the \ion{Fe}{1} 3280.666~\AA\ line ($\Eexc$ = 3.283~eV, log~$gf = -2.35$), which was predicted by R. Kurucz and presented in the VALD database. The \ion{Fe}{1} line forms in the transition between the observed levels \eu{z}{5}{D}{\circ}{1} and \eu{5d}{5}{F}{}{2}, and thus its wavelength is reliable.
 In its red wing, the \ion{Ag}{1} line overlaps with \ion{Zr}{2} 3280.735~\AA\ and \ion{Mn}{1} 3280.76~\AA. In order to reproduce the most pronounced absorption features in the 3280.2--3281.0~\AA\ spectral range, we were forced to adjust abundances of N, Mn, and Fe, by +0.17, +0.22, and $-0.1$~dex, respectively, compared to the solar abundances of \citet{Lodders2021}. 
Absorptions of unknown origin are present in the blue wing of the \ion{Ag}{1} line, and we are unable to estimate their influence on the derived Ag abundance.

The Ag abundances were derived under the LTE assumption and in the two NLTE line formation scenarios, with applying the YV26 and CA26 collisional data sets, see Table~\ref{tab:sun}. Despite consistent abundances from the two lines, hereafter, 
 we refer to the results based on \ion{Ag}{1} 3382~\AA, only. The LTE abundance is lower than the meteoritic one, $\eps{met}$ = 1.21$\pm$0.02 \citep{Lodders2021}, by 0.32~dex.
NLTE leads to weakened \ion{Ag}{1} lines and the higher derived abundance, such that the discrepancy with the meteoritic one reduced to $-0.11$~dex and $-0.06$~dex in the variants CA26 and YV26, respectively. 
Implementing $gf$(JS26) in the model atom instead of $gf$(K11) leads to slightly higher CA26 and YV26 abundances, by 0.02~dex and 0.03~dex, respectively.
Greater shifts in the derived Ag abundance, of up to 0.07~dex, can be due to using different $gf$(\ion{Ag}{1} 3382~\AA) in the synthetic spectra calculations, namely log~$gf = -0.334$ \citep{1990ZPhyD..16...87C}, or $-0.356$ \citep{2026A&A...711A.155C}, or $-0.4$ \citep{1980wtpa.book.....W}.

{\it Comparison with the literature data.} \citet{2026A&A...711A.155C} calculated the solar LTE and NLTE abundances of Ag using the 3D and 1D model atmospheres. For a common $gf$-value, our LTE abundance from \ion{Ag}{1} 3382~\AA\ is slightly higher, by 0.02~dex, compared to that in the 1D model of \citet{2026A&A...711A.155C}. In the CA26 scenario, the NLTE abundance correction ($\Delta_{\rm NLTE} = \eps{NLTE} - \eps{LTE}$) for \ion{Ag}{1} 3382~\AA\ agrees within 0.01~dex with that of \citet{2026A&A...711A.155C} in the 1D model. The same authors found that neglecting the 3D effects leads to an underestimation of the NLTE abundance derived from \ion{Ag}{1} 3382~\AA, by 0.06~dex. Their recommended 3D NLTE abundance $\eps{Ag}$ = 1.15$\pm$0.08 is lower than the meteoritic one, although the difference is within the error bars.  

\section{Stellar abundances of Ag} \label{Sect:abund}

\subsection{Stellar sample, observations, atmospheric parameters} \label{Sect:sample}

This study aims to derive the Ag abundances for the same stellar sample, for which 
\citet[][MS2025]{pd_nlte} calculated the NLTE abundances of Pd.
 We used the same observed spectra taken from the spectral archives of the ESO instruments: the Ultraviolet and Visual Echelle Spectrograph (UVES) and the High Accuracy Radial velocity Planet Searcher (HARPS). The \ion{Ag}{1} 3382~\AA\ line was reliably detected in 34 stars, which cover the $-2.20 \le$ [Fe/H] $\le -0.56$ metallicity range. 
 The 32 of them have 0.3 $\le$ [Eu/Fe] $\le$ 1 and [Ba/Eu] $< 0$, according to MS2025, and are related to the r-process enhanced stars of the category r-I introduced by \citet{HERESI}. The remaining two stars, HD~195633 and HD~204543, have [Eu/Fe] $\sim$ 0.2 and thus a notable contribution of the r-process to their heavy element abundances.
 Stellar sample of MS2025 includes two stars, which are poor in the r-process elements and reveal different Pd origin compared to the r-I stars. It would be important to check silver in these stars. However, the Ag abundance cannot be constrained for CD~-24~1782 due to bad spectrum around 3382~\AA. The second star, HD~122563, was included in our sample with 
 an upper limit for the Ag abundance. 
  The selected stars are listed in Table~\ref{tab:abundances} together with their atmospheric parameters taken from MS2025.
We used the model atmospheres obtained by MS2025 via the interpolation in the MARCS grid for given $\Teff$, \logg, and [Fe/H].

In order to improve the statistics in the [Fe/H] $< -2$ range, we included the four giants  from \citet[][LB2022]{Lombardo2022}. Their spectra were taken from the UVES archive and the $\Teff$, \logg\ values from LB2022. The Fe abundances were revised, using the same \ion{Fe}{2} lines and $gf$-values, as in MS2025. The empirical relation from \citet{2025A&A...699A.262S} was
 applied for computing $\xi_t$.

For testing the r-process models, it is important to measure the Ag abundance of the known r-II stars.
In this study, we use two r-II stars, CS22892-052 \citep{Sneden2003} and CS31082-001 \citep{hill2002}. Their spectra were taken from the UVES archive and the atmospheric parameters $\Teff$, \logg, $\xi_t$ from \citet[][MR2025]{2025AJ....169..172M}. We revised the [Fe/H] values, as it was done for the LB2022 stars.

Stellar sample of MS2025 has six common stars with LB2022 and six common stars with MR2025.
The mean differences between MS2025 and LB2022, $\Delta \Teff = -27\pm$67~K and $\Delta$\logg\ = 0.06$\pm$0.14, are well below the uncertainties of 100~K and 0.1~dex, as obtained by  MS2025.	
 There is a perfect agreement between MS2025 and MR2025 in $\Teff$ (within 18~K) and \logg\ (within 0.06~dex) for two dwarfs. For the four giants, the mean differences amount to $-118\pm$29~K for $\Teff$ and 0.08$\pm$0.17~dex for \logg. 

\begin{deluxetable*}{lcccccrrrrcccc}
\tablewidth{0pt}
\tablecaption{Atmospheric parameters for sample stars and LTE and NLTE abundances of Ag and Eu  \label{tab:abundances}}
\tablehead{
\multicolumn{1}{c}{Star} & \colhead{$\Teff$,} & \colhead{\logg} & \colhead{[Fe/H]} & \colhead{$\xi_t$} & Ref & \colhead{[N/Fe]} & \multicolumn{3}{c}{$\eps{Ag}$} & \colhead{$\Delta_{\rm NLTE}$} & \colhead{[Ag/Fe]} & \multicolumn{2}{c}{[Eu/Fe]} \\
  \cline{8-10}
  \cline{13-14}
 & \multicolumn{1}{c}{K} & & & & & \colhead{LTE} & \colhead{LTE} & \colhead{CA26} & \colhead{YV26} &  \colhead{YV26} & \colhead{YV26} & \colhead{LTE} & \colhead{NLTE}
}
\startdata 
HD 2796  &  4960 & 1.83 & --2.33 & 1.9 & 1 & --0.18 & --1.21 & --0.53 & --0.49 & 0.72 & 0.63 & 0.10 & 0.28 \\
HD 3567  & 6180 & 4.06 & --1.14 & 1.4 & 2 & --0.05 &   0.47 &   0.87 &   0.92 &  0.45 &  0.85 & 0.57 & 0.68 \\
HD 5426  &  5056 & 2.45 & --2.24 & 1.5 & 1 & --0.19 & --0.95 & --0.32 & --0.27 & 0.68 & 0.76 & 0.69 & 0.81 \\
HD 8724  & 4560 & 1.29 & --1.71 & 1.5 & 2 & --0.23 & --1.10 & --0.33 & --0.29 &  0.81 &  0.21 & 0.20 & 0.34 \\
HD 22879 & 5940 & 4.28 & --0.89 & 1.2 & 2 & --0.22 &   0.27 &   0.65 &   0.71 &  0.44 &  0.39 & 0.38 & 0.45 \\
HD 25704 & 5940 & 4.23 & --0.88 & 1.2 & 2 & --0.06 &   0.27 &   0.65 &   0.71 &  0.44 &  0.38 & 0.28 & 0.36 \\
HD 29907 & 5500 & 4.59 & --1.51 & 0.6 & 2 & --0.53 &   0.08 &   0.40 &   0.58 &  0.50 &  0.88 & 0.64 & 0.70 \\
HD 59392 & 6045 & 3.95 & --1.57 & 1.4 & 2 & --0.10 & --0.13 &   0.31 &   0.37 &  0.50 &  0.73 & 0.50 & 0.68 \\
HD 60319 & 6170 & 4.09 & --0.86 & 1.4 & 2 &   0.10 &   0.45 &   0.82 &   0.87 &  0.42 &  0.52 & 0.35 & 0.45 \\
HD 63077 & 5870 & 4.19 & --0.82 & 1.2 & 2 & --0.17 &   0.39 &   0.77 &   0.84 &  0.45 &  0.45 & 0.32 & 0.39 \\
HD 63598 & 5900 & 4.21 & --0.87 & 1.2 & 2 & --0.26 &   0.34 &   0.72 &   0.78 &  0.44 &  0.44 & 0.31 & 0.39 \\
HD 76932 & 6000 & 4.14 & --0.88 & 1.3 & 2 & --0.16 &   0.41 &   0.79 &   0.85 &  0.44 &  0.52 & 0.39 & 0.47 \\
HD 83212 & 4550 & 1.00 & --1.44 & 1.9 & 2 & --0.02 & --0.46 &   0.16 &   0.20 &  0.66 &  0.43 & 0.28 & 0.37 \\
HD 97320 & 6110 & 4.26 & --1.12 & 1.4 & 2 & --0.21 &   0.21 &   0.61 &   0.66 &  0.45 &  0.57 & 0.26 & 0.37 \\
HD 102200 & 6115 & 4.29 & --1.13 & 1.4 & 2 & --0.41 &   0.06 &   0.46 &   0.51 &  0.45 &  0.43 & 0.43 & 0.53 \\
HD 103723 & 6180 & 4.18 & --0.81 & 1.4 & 2 & --0.20 &   0.51 &   0.88 &   0.93 &  0.42 &  0.53 & 0.41 & 0.50 \\
HD 105004 & 5880 & 4.40 & --0.79 & 1.1 & 2 & --0.31 &   0.31 &   0.68 &   0.75 &  0.44 &  0.33 & 0.25 & 0.32 \\
HD 106038 & 6115 & 4.29 & --1.34 & 1.3 & 2 & --0.16 &   0.14 &   0.55 &   0.61 &  0.47 &  0.74 & 0.34 & 0.45 \\
HD 108317 & 5270 & 2.96 & --2.13 & 1.2 & 2 &        & --1.06 & --0.49 & --0.43 &  0.63 &  0.49 & 0.44 & 0.56 \\
HD 108577 & 5089 & 2.04 & --2.13 & 1.8 & 1 & --0.26 & --0.91 & --0.26 & --0.22 & 0.69 & 0.70 & 0.46 & 0.61 \\
HD 111980 & 5910 & 4.00 & --1.09 & 1.3 & 2 & --0.36 &   0.14 &   0.56 &   0.62 &  0.48 &  0.50 & 0.28 & 0.38 \\
HD 113679 & 5680 & 4.02 & --0.64 & 1.2 & 2 & --0.46 &   0.38 &   0.76 &   0.82 &  0.44 &  0.25 & 0.26 & 0.31 \\
HD 120559 & 5450 & 4.43 & --0.98 & 0.9 & 2 & --0.19 &   0.16 &   0.47 &   0.61 &  0.45 &  0.38 & 0.37 & 0.42 \\
HD 121004 & 5910 & 4.61 & --0.66 & 1.1 & 2 & --0.24 &   0.61 &   0.93 &   1.03 &  0.42 &  0.48 & 0.46 & 0.51 \\
HD 122563 & 4600 & 1.40 & --2.46 & 1.6 & 2 &   0.03 &$<$--2.14 &$<$--1.29 & $<$--1.26 &  0.88 &$<$--0.01 & --0.72 & --0.50 \\
HD 126681 & 5640 & 4.65 & --1.13 & 0.9 & 2 & --0.35 &   0.13 &   0.47 &   0.60 &  0.47 &  0.52 & 0.46 & 0.52 \\
HD 132475 & 5810 & 3.86 & --1.41 & 1.3 & 2 & --0.17 & --0.27 &   0.18 &   0.24 &  0.51 &  0.44 & 0.29 & 0.40 \\
HD 165195 & 4420 & 0.97 & --2.15 & 2.0 & 2 &   0.08 & --1.46 & --0.57 & --0.54 &  0.92 &  0.40 & 0.30 & 0.45 \\
HD 175179 & 5850 & 4.30 & --0.71 & 1.2 & 2 & --0.08 &   0.46 &   0.82 &   0.89 &  0.43 &  0.39 & 0.32 & 0.38 \\
HD 188510 & 5560 & 4.68 & --1.51 & 0.8 & 2 & --0.44 & --0.41 & --0.07 &   0.07 &  0.48 &  0.37 & 0.41 & 0.47 \\
HD 189558 & 5765 & 3.87 & --1.12 & 1.3 & 2 & --0.41 &   0.14 &   0.58 &   0.64 &  0.50 &  0.55 & 0.35 & 0.44 \\
HD 195633 & 6130 & 3.92 & --0.56 & 1.5 & 2 &   0.03 &   0.52 &   0.87 &   0.91 &  0.39 &  0.26 & 0.12 & 0.21 \\
HD 204543 & 4670 & 1.41 & --1.59 & 1.9 & 2 &   0.04 & --1.11 & --0.41 & --0.39 &  0.72 & --0.01 & 0.04 & 0.17 \\
HD 205650 & 6080 & 4.40 & --1.17 & 1.2 & 2 &   0.13 &   0.19 &   0.58 &   0.64 &  0.45 &  0.60 & 0.50 & 0.61 \\
CD --33 3337 & 6000 & 3.88 & --1.34 &1.4 & 2 & --0.06 & --0.19 &   0.25 &   0.30 &  0.49 &  0.43 & 0.25 & 0.37 \\
CD --57 1633 & 6050 & 4.36 & --0.88 &1.2 & 2 & --0.36 &   0.48 &   0.84 &   0.91 &  0.43 &  0.58 & 0.42 & 0.50 \\
BD +05 3098 &  5013 & 2.30 & --2.45 & 1.6 & 1 & --0.21 & --1.07 & --0.43 & --0.39 & 0.68 & 0.85 & 0.37 & 0.52 \\
BD +20 0571 & 6130 & 4.05 & --0.91 & 1.4 & 2&   0.25 &   0.69 &   1.06 &   1.11 &  0.42 &  0.81 &      &      \\
BD --01 2582 & 5220 & 2.82 & --2.20 &1.6 & 2& --0.37 & --0.96 & --0.37 & --0.31 &  0.65 &  0.68 & 0.68 & 0.80 \\
CS22892-052 & 4965 & 1.90 & --2.74 &1.6 & 3 & --0.62 & --0.69 & --0.03 &   0.02 &  0.71 &  1.55 & 1.47 & 1.61 \\
CS31082-001 & 5010 & 2.13 & --2.54 &1.65& 3 &  0.64 & --0.57 &   0.06 &   0.13 &  0.70 &  1.46 & 1.61 & 1.70 \\
\enddata
\tablecomments{ $\xi_t$ in \kms . Ref: 1 = \citet{Lombardo2022}, 2 = MS2025, 3 = \citet{2025AJ....169..172M}.}
\end{deluxetable*}

\subsection{Silver abundances}

Stellar abundances were derived from \ion{Ag}{1} 3382~\AA, with using the code \textsc{synthV\_NLTE} \citep{2019ASPC}, which computes the synthetic spectra, and the IDL visualization program \textsc{BinMag} \citep{2018ascl.soft05015K}, which enables us to obtain the best fit to the observed spectrum. The line list and required atomic parameters were from VALD. 
 Everywhere, we adopted the solar isotopic ratio $^{107}$Ag : $^{109}$Ag = 51.84 : 48.16 \citep{Lodders2021}. Using the alternative value, that is the r-process ratio 56.6 : 43.4 \citep{Prantzos2020}, leads to an abundance shift of no more than 0.01~dex. 

\begin{figure}   
	\plotone{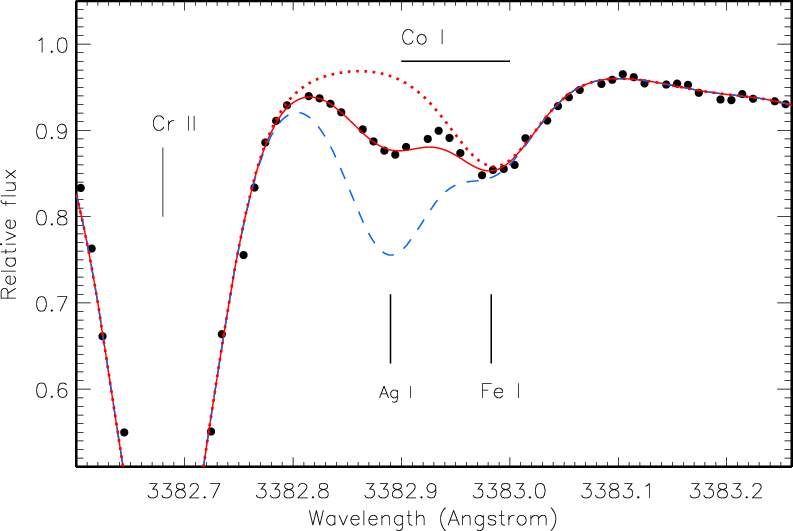}
	\plotone{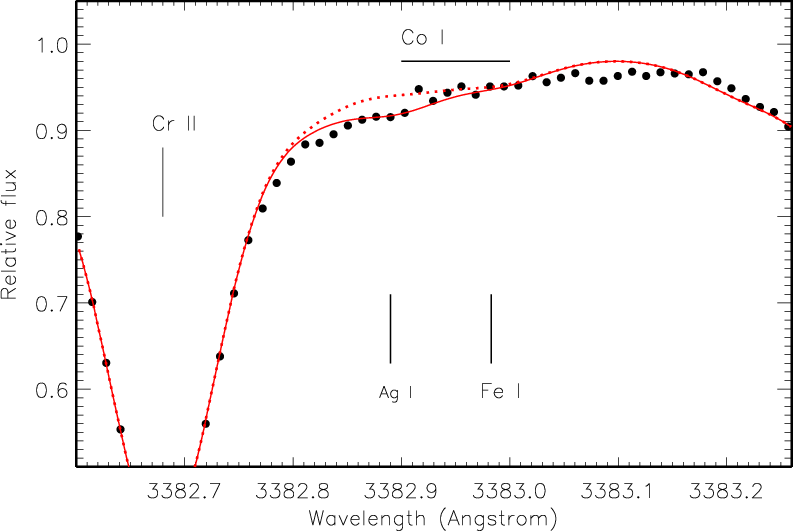}
	\caption{\ion{Ag}{1} 3382~\AA\ line in spectra (bold dots) of two stars. 
		Top panel: best NLTE(YV26) fit (continuous curve) for HD~189558. The same abundance was used to compute the LTE profile (dashed curve). 
	Bottom panel: continuous curve is the theoretical spectrum corresponding to the Ag abundance upper limit for HD~122563.
		In each panel, the synthetic spectrum without the contribution of the \ion{Ag}{1} line is shown by the dotted curve. The NLTE abundances we obtained are indicated in Table~\ref{tab:abundances}. 
		\label{fig:hd189558}}
\end{figure}

The Ag abundance was derived if the line core of \ion{Ag}{1} 3382~\AA\ is clearly seen in the observed spectrum. Quality of spectrum fitting for one of the sample stars, HD~189558, is illustrated in Fig.~\ref{fig:hd189558}. For comparison, the bottom panel of Fig.~\ref{fig:hd189558} shows the theoretical spectrum corresponding to the Ag abundance upper limit for HD~122563. The background opacity for \ion{Ag}{1} 3382~\AA\ is contributed from the \ion{Cr}{2}, \ion{Fe}{1}, \ion{Co}{1} lines, which are indicated in Fig.~\ref{fig:hd189558}, the far wing of \ion{Ti}{2} 3383.76~\AA, and the molecular NH 3382.818, 3382.900, 3382.926~\AA\ lines. Effect of the NH lines is minor for the \logg\ $> 3$ stars (dwarfs), but notable for the cool giant stars (\logg\ $\le 3$). For example, in the 4420/0.97/$-2.15$ model, a reduction of the N abundance by 0.2~dex leads to the higher continuous flux around \ion{Ag}{1} 3382~\AA\ by 2\%, which translates to the increased Ag abundance, by 0.05~dex. The N abundance necessary for analysis of the 3383~\AA\ blend was determined for each star 
 from the NH band at 3360~\AA\ (see Table~\ref{tab:abundances} for the [N/Fe] values). For best fitting the \ion{Ag}{1} line, it was necessary to adjust abundances of Ti, Cr, Fe, and Co. For the majority of stars, the changes in abundances did not exceed 0.1~dex compared to the metallicity scaled solar abundances. In the case of HD~189558, only Fe abundance needed to be changed, by +0.12~dex. 

Our results for three line-formation scenarios: LTE, NLTE(CA26), and NLTE(YV26) are presented in Table~\ref{tab:abundances}. The [Ag/Fe] values were computed from the stellar NLTE(YV26) abundances and the Solar System abundance $\eps{Ag,met}$ = 1.21 \citep{Lodders2021}. 
NLTE leads to weakened \ion{Ag}{1} lines and higher abundances compared to the LTE case. The NLTE corrections are larger for the giants, with $\Delta_{\rm NLTE}$ = 0.63 to 0.92~dex (YV26 scenario), than for the dwarfs, with $\Delta_{\rm NLTE}$ = 0.39 to 0.51~dex. 
The NLTE effects are slightly weaker in the CA26 scenario, with the YV26 -- CA26 abundance difference reducing toward lower \logg\ from 0.18~dex in the 5500/4.59/$-1.51$/[Ag/Fe] = 0.9 model down to 0.02~dex in the 4670/1.41/$-1.59$/[Ag/Fe] = 0 model.

\subsection{Uncertainties in the Ag NLTE abundances}

\begin{deluxetable}{lcc}
	\tablewidth{0pt}
	\tablecaption{Error estimates (dex) for the abundances derived from \ion{Ag}{1} 3382~\AA\ in HD~165195 and HD~3567 \label{tab:uncertainties}}
	\tablehead{
		& \colhead{4420/0.97/$-2.15$} & \colhead{6180/4.06/$-1.14$} \\
		& \colhead{$\eps{Ag}$ = $-0.54$} & \colhead{$\eps{Ag}$ = 0.92} 
	}	
	\startdata
	& \multicolumn{2}{c}{Changes relative to $\eps{Ag}$} \\
	\multicolumn{3}{l}{Line formation treatment} \\
	LTE  & $-0.92$ & $-0.45$  \\
	CA26, $\sigma_{\rm col}$ & $-0.03$ & $-0.05$ \\
	$gf$(JS26), $\sigma_{gf}$ & $-0.06$ & $-0.08$ \\
	\multicolumn{3}{l}{Atmospheric parameters} \\
	$\Teff$: +100~K, $\sigma_{\rm Teff}$ &  ~~0.19 & ~~0.12  \\
	\logg : +0.1, $\sigma_{\rm logg}$   & ~~0.02 & $-0.01$ \\
	& \multicolumn{2}{c}{Abundance errors	} \\
	$\sigma_{\rm Ag}$ & ~~0.05 & ~~0.05 \\
	$\sigma_{\rm Fe}$ & ~~0.05 & ~~0.04 \\
	$\sigma_{\rm tot}$ & ~~0.21 & ~~0.16 \\
	\enddata
\end{deluxetable}

The uncertainty in the NLTE abundance from \ion{Ag}{1} 3382~\AA\ is contributed from the observational errors ($\sigma_{\rm Ag}$) due to the uncertainties in the continuum placement and accounting for the blending lines, from the uncertainties in NLTE treatment ($\sigma_{\rm NLTE}$) and atmospheric parameters.
Our calculations showed that, for the majority of sample stars, $\sigma_{\rm Ag}$ cannot be larger than 0.05~dex. For $\sigma_{\rm NLTE}$, we adopted the quadratic sum of the abundance difference between the variants YV26 and CA26 ($\sigma_{\rm col}$) and the abundance shift due to replacing $gf$(K11) with $gf$(JS26) in the SE calculations ($\sigma_{gf}$). 
 Variations in the photoionization cross-sections for \ion{Ag}{1} lead to negligibly small shifts in the derived abundances, as shown first by \citet{2026A&A...711A.155C} for the Sun and checked in this study for several stars.
For the uncertainties in $\Teff$ and \logg, we adopted $\pm$100~K and $\pm$0.1~dex, as discussed by MS2025. The \ion{Ag}{1} lines are weak in our stellar sample, and variations of 0.2~\kms\ in $\xi_t$ produce the Ag abundance shifts of less than 0.01~dex.

Table~\ref{tab:uncertainties} summarizes the abundance errors for the coolest and hottest stars of the sample. It can be seen that the uncertainty in $\Teff$ is the greatest source of the Ag abundance error. 
The total uncertainty $\sigma_{\rm tot}$ in [Ag/Fe] was computed for each star by the quadratic sum of $\sigma_{\rm Ag}$, $\sigma_{\rm NLTE}$, abundance errors due to the uncertainties in $\Teff$ and \logg, and the stochastic error of the Fe abundance $\sigma_{\rm Fe}$: 
$$\sigma_{\rm tot} = \sqrt{\sigma_{\rm Ag}^2 + \sigma_{\rm Fe}^2 + \sigma_{\rm Teff}^2 + \sigma_{\rm logg}^2 + \sigma_{\rm NLTE}^2}.$$

\subsection{Europium abundances}

\citet{nlte_eu2026} updated the NLTE method for \ion{Eu}{2} by implementing accurate rate coefficients for excitation and ion pair production in the \ion{Eu}{2} + \ion{H}{1} collisions.
This motivated us to revise the Eu NLTE abundances from MS2025, which were based on an approximate treatment of hydrogenic collisions with classical Drawinian rates \citep[computed according to][]{Steenbock1984}. As in  MS2025, the Eu abundance was derived from \ion{Eu}{2} 4129~\AA. Due to multiple HFS components, this line is insensitive to variations in $\xi_t$, even being strong in the r-II stars. For example, in the 4965/1.90/$-2.74$/[Eu/Fe] = 1.61 model, the abundance difference between using $\xi_t$ = 1.6 and 1.8~\kms\ amounts to 0.01~dex.

For the majority of sample stars, the abundance difference between this study (Table~\ref{tab:abundances}) and  MS2025 does not exceed 0.02~dex. A maximal difference of 0.08~dex was obtained for the 4560/1.29/$-1.71$/[Eu/Fe] = 0.3 model. 

\section{Galactic trends of the elemental ratios} \label{Sect:trends}

\begin{figure}  
	\plotone{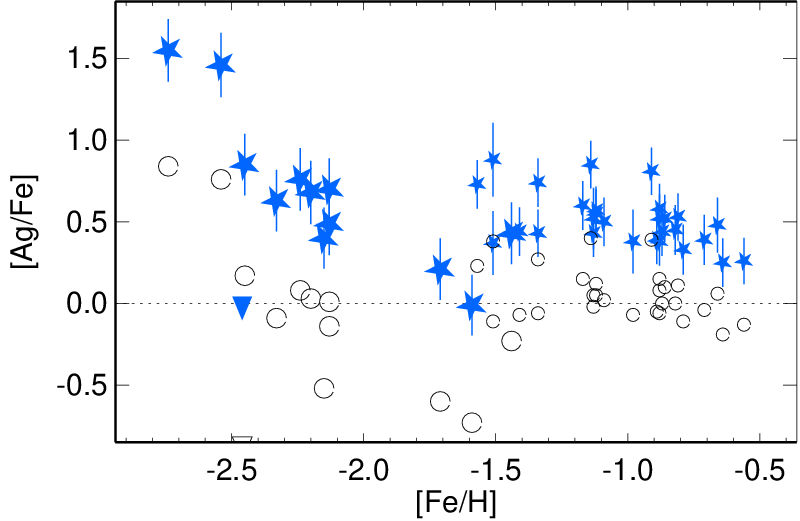}
	\plotone{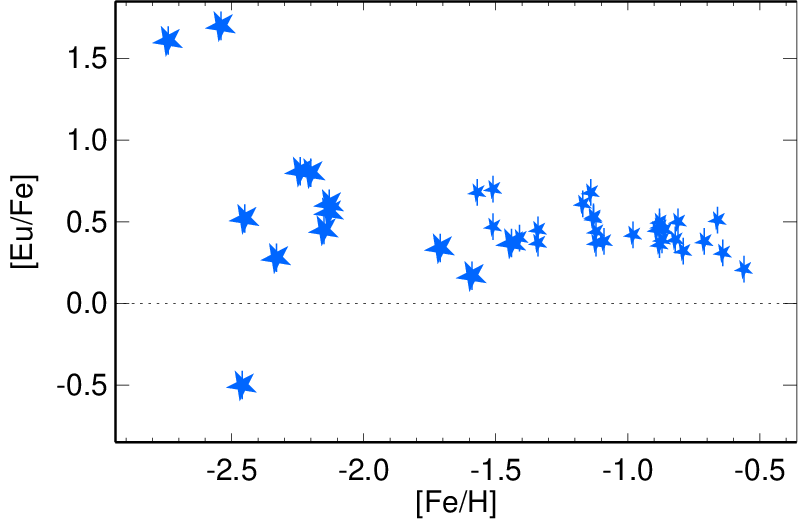}
	\caption{Top panel: [Ag/Fe] NLTE (YV26, stars) and LTE (open circles) abundance ratios for our sample stars. The filled and open inverted triangles show the upper limits for the NLTE and LTE abundances of HD~122563. The smaller and larger symbols correspond to the \logg\ $> 3.0$ and \logg\ $\le 3.0$ stars, respectively. For each star, the error bar corresponds to  $\sigma_{\rm tot}$. Bottom panel: [Eu/Fe] NLTE abundance ratios for the same stars.
		\label{fig:abundances}}
\end{figure}

In NLTE, independent of the variants YV26 or CA26, our sample stars, except  HD~122563 (known r-poor star) and HD~204543, are enhanced with silver relative to iron (Fig.~\ref{fig:abundances}, top panel). The strongest enhancements, with [Ag/Fe] $\simeq$ 1.5, are observed for the r-II stars. 
In the remaining giants, the [Ag/Fe] values approach the mean for the dwarf sample, $\simeq$0.5. From Fig.~\ref{fig:abundances}, it is evident that the Galactic trends of the NLTE [Ag/Fe] and [Eu/Fe] ratios are very similar, suggesting a common origin of Ag and Eu. 

\begin{figure}  
	\plotone{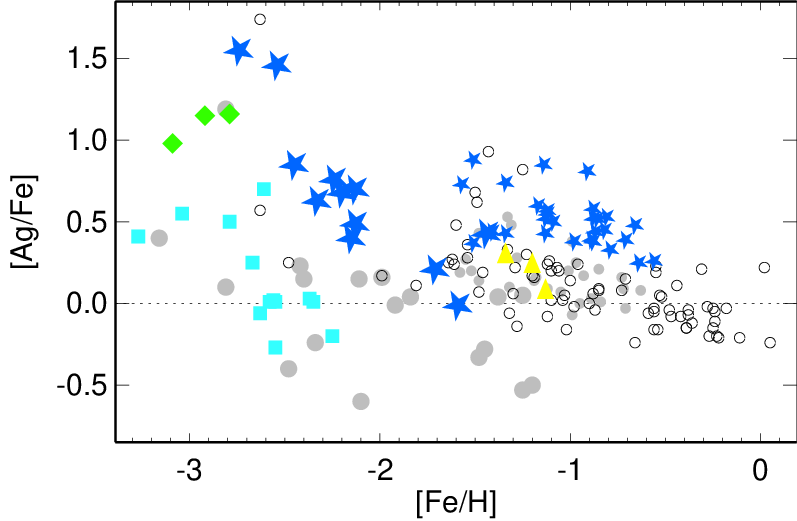}
	\caption{NLTE(YV26) abundance ratios [Ag/Fe] of our sample stars (stars) compared to the  LTE values from \citet[][filled gray circles]{2012A&A...545A..31H}, \citet[][open circles]{2015A&A...579A...8W}, \citet[][cyan squares]{2025ApJ...991..112H}, \citet[][yellow triangles]{2024A&A...690A.262B}.
 The green rhombi show three r-II stars from \citet{Sneden2003} and \citet{hill2002,2017A&A...607A..91H}.		
		 The smaller and larger symbols correspond to the dwarfs and giants, respectively. 
		\label{fig:literature}}
\end{figure}

The conclusion is different in the LTE case. 
Our LTE [Ag/Fe] ratios (Fig.~\ref{fig:abundances}) and the data available in the literature (Fig.~\ref{fig:literature}) reveal close-to-solar values for the dwarfs, which occupy the [Fe/H] $> -1.5$ region, and a scatter for the giants at the lower metallicities. Such a behavior resembles that for [Sr/Fe] and [Ba/Fe] (for example, Fig.~7 in MS2025), suggesting common mechanisms and astrophysical sites for production of Ag and Sr, Ba.

\begin{figure*}  
\plottwo{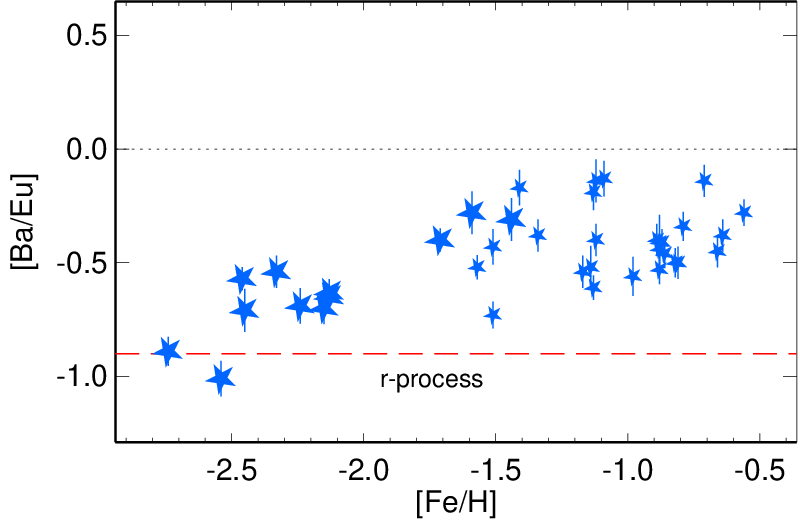}{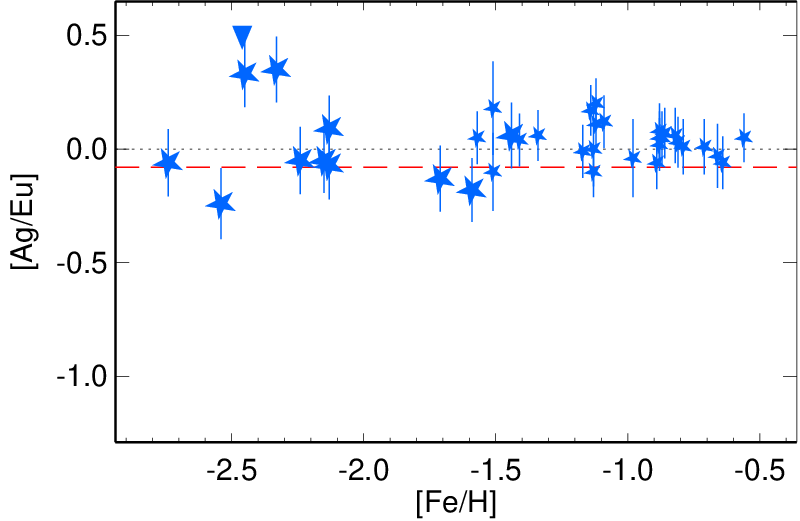}
\plottwo{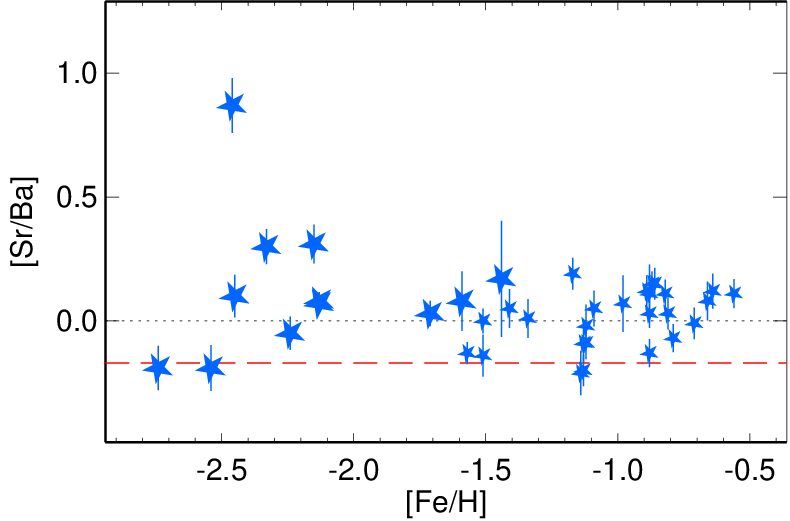}{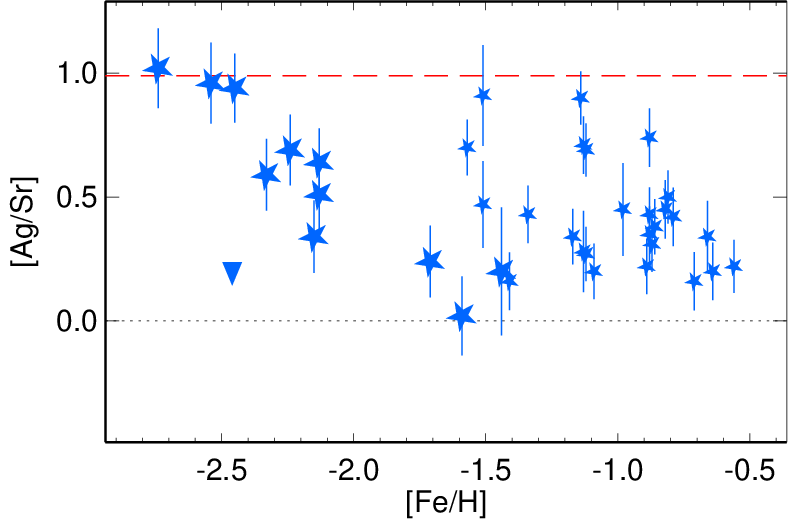}
\plottwo{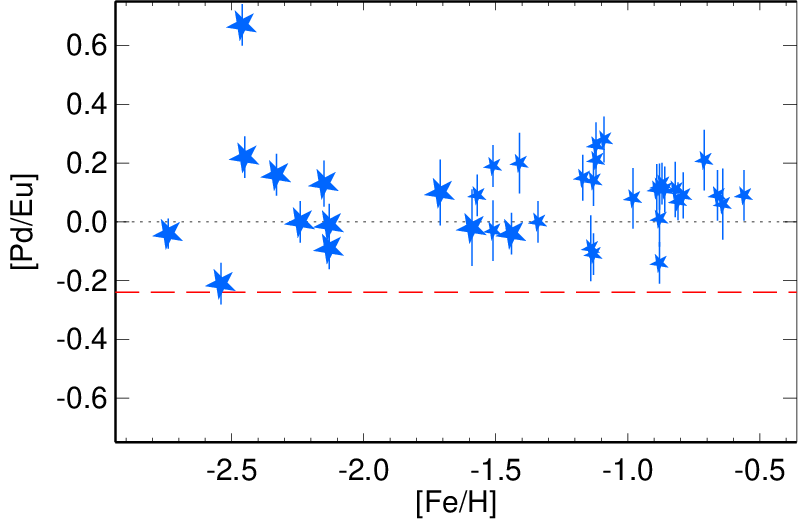}{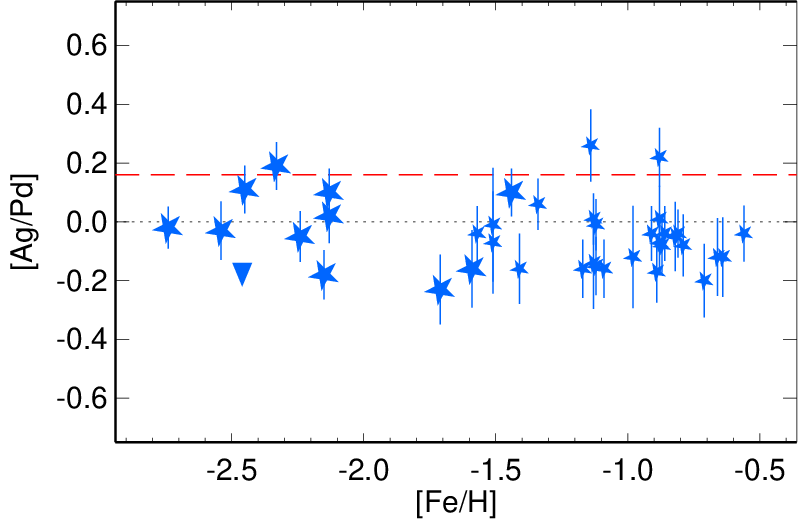}
\caption{NLTE abundance ratios among Ag, Sr, Ba, Eu, and Pd for our sample stars. In each panel, the dashed red line indicates the  r-process ratios as predicted by \citet{Prantzos2020}. Symbols are the same as in Fig.~\ref{fig:abundances}.
\label{fig:ratios}}
\end{figure*}

\begin{deluxetable}{cccc}
	\tablewidth{0pt}
	\tablecaption{NLTE abundance ratios of the r-II stars compared to the solar r-residuals from \citet{Prantzos2020} \label{tab:rII}}
	\tablehead{
	& \colhead{CS22892-052} & \colhead{CS31082-001} & \colhead{r-process} 
	}	
	\startdata
 \ [Ba/Eu] & $-0.89$ & $-1.01$ &  $-0.90$ \\
 \ [Sr/Ba] & $-0.19$ & $-0.19$ &  $-0.17$ \\
 \ [Ag/Eu] & $-0.06$ & $-0.24$ &  $-0.08$ \\
 \ [Ag/Ba] & ~~0.83  & ~~0.77  &  ~~0.82 \\
 \ [Ag/Sr] & ~~1.02  & ~~0.96  &  ~~0.99 \\
 \ [Ag/Pd] & $-0.02$ & $-0.03$ &  ~~0.16 \\
 \ [Pd/Eu] & $-0.04$ & $-0.21$ &  $-0.24$ 	
	\enddata
\end{deluxetable}

Using the NLTE abundances derived in this study and by MS2025, we plotted in Fig.~\ref{fig:ratios} the 
abundance ratios between Ag and the best representatives of the elements produced effectively in the weak s-process (Sr), the main s-process (Ba), and the r-process (Eu).
Following MS2025, we excluded a CEMP-r/s star BD-01~2582 and a Be-rich star HD~106038 from analysis of the Galactic chemical evolution.

 In Fig.~\ref{fig:ratios}, the observations are confronted to the r-process model. It represents the solar r-residuals based on the s-process calculations of \citet{Prantzos2020}, which included contributions from intermediate-mass stars (main s-process) and massive rotating stars (weak s-process).
We found that the predicted r-process ratios among Eu, Ba, Sr, and Ag agree very well with the observations of two benchmark r-II stars (Table~\ref{tab:rII}). 
For the ratios involving Pd, there are differences at the level of 0.2~dex.
 They cannot be entirely explained by the uncertainties in the r-process model, because the [Pd/Eu] ratios of CS31082-001 and also the r-II stars HE2327-5642 and HE1219-0312 from MS2025 agree within 0.1~dex with the predicted r-process one.
 
 With the revised Eu abundances, we confirm that the [Ba/Eu] ratios of the [Fe/H] $\gtrsim -2.4$ stars are higher compared to the r-process one and increase toward higher metallicity. This means that the main s-process in AGB stars operated already, when the Galactic Fe abundance reached [Fe/H] $\simeq -2.4$, and its contribution to the Ba production increased with time. 
 
In contrast to the [Ba/Eu] behavior and in line with a prediction of relatively small contribution of the s-process to the solar Ag abundance \citep[20.7\%,][]{Prantzos2020}, the [Ag/Eu] trend is nearly flat. In the [Fe/H] $\lesssim -1.6$ stars, [Ag/Eu] is close to the r-process ratio of $-0.08$. The exceptions are HD~2796 and BD~+05 3098, with [Ag/Eu] = 0.35  and 0.33, and HD~122563, with [Ag/Eu] $< 0.5$.
The higher metallicity stars reveal close-to-solar values. 

The [Ag/Sr] ratios in our sample stars, except the r-II stars, are lower than the r-process one due to a contribution of the s-process to Sr production. However, the data reveal a scatter rather than an expected decreasing trend. Silver is tightly correlated with Pd throughout the whole metallicity range we investigated, suggesting their common origin.

To summarize, the NLTE analysis of the \ion{Ag}{1} lines in a sample of MP stars provides evidence for a dominant contribution of the main r-process to Galactic silver throughout the whole metallicity  range we investigated, with a contribution of the main s-process at the level of 20\%\ for the [Fe/H] $\gtrsim -1.6$ epoch.  

 The clear exception is HD~122563, in line with the literature studies. In their LTE analysis, \citet{2017ApJ...837....8A} found the decrease in the abundances between Mo and Pd and argued that it can be attributed to the weak r-process. 
  \citet{2024ApJ...966...11P} examined models of the weak r-process and $\nu p$-process occurring in early core-collapse supernovae explosions
  and concluded that different astrophysical conditions are required for fitting observations of different elements in HD~122563: Sr-Zr, Mo, Ru and Pd. It would be important to reexamine their nucleosynthetic models with the NLTE abundances. 
\citet{2025A&A...699A.262S} provided evidence that the Sr abundance of HD~122563 was contributed at the level of 45~\%\ from the s-process in extremely metal-poor ([Fe/H] $< -3$) fast-rotating massive stars \citep{2008ApJ...687L..95P,2014A&A...565A..51C}.  
 Based on the NLTE analysis, \citet{pd_nlte} proposed that the early s-process contributed also to Pd. 
The upper limit for the Ag abundance, we obtained in this study, suggests an enhancement of Ag relatively to Eu, and it can be explained by a contribution of the early s-process. Calculations of \citet{2016MNRAS.456.1803F} for the rotating (40\%\ of
the critical velocity) model of 25~$M_\odot$ with [Fe/H] = $-3.8$ showed that the production factor for Ag is greater than that in the nonrotating model, by about 2.5~dex.
  There is a hint that an additional source of heavy elements besides the main r-process is required for VMP stars HD~2796 and BD~+05 3098, which reveal enhancements of Ag relative to Eu, although of less magnitude than that for HD~122563.

\section{Conclusions}\label{Sect:conclusion}

In this study, we present the new NLTE method for determinations of stellar Ag abundances. 
A comprehensive model atom of \ion{Ag}{1} was built from the radiative and collisional data available in the literature, including  the rate coefficients for inelastic processes in the \ion{Ag}{1} + \ion{H}{1} collisions from calculations of \citet{2026A&A...711A.155C} and \citet{ag_hyd}. 

As a first application of the new model atom, we analyzed two \ion{Ag}{1} lines in the solar disk center intensity spectrum \citep{Delbouille1973} using the synthetic spectrum method with the classical MARCS model atmosphere. The NLTE abundances we obtained in the variants YV26 and CA26 are $\eps{\odot,Ag}$ = 1.15 and 1.10, respectively. Consistency with the meteoritic abundance, $\eps{met,Ag}$ = 1.21 \citep{Lodders2021}, can be achieved after applying the 3D correction of +0.06~dex from \citet{2026A&A...711A.155C}. The mean LTE abundance is lower than the meteoritic one, by 0.35~dex. 

The NLTE abundances of Ag, in both YV26 and CA26 variants, were determined for the first time for 41 stars in the $-2.74 \le$ [Fe/H] $\le -0.56$ metallicity range. They are based on high-resolution spectra from the ESO archives and the synthetic spectrum method with the MARCS model atmospheres. The NLTE abundance corrections for \ion{Ag}{1} 3382~\AA\ are larger for the giants, with $\Delta_{\rm NLTE}$(YV26) = 0.63~dex up to 0.92~dex, than the dwarfs, where $\Delta_{\rm NLTE}$(YV26) = 0.39~dex up to 0.51~dex. 

In the NLTE analysis, independent of the YV26 or CA26 variants, we found that
\begin{itemize}
	\item our sample stars, except HD~204543 and HD~122563, are all enhanced with Ag relatively to Fe and form the Galactic trend of [Ag/Fe], which is very similar to that for [Eu/Fe];
	\item the [Ag/Eu], [Ag/Ba], and [Ag/Sr] abundance ratios of the r-II stars CS22892-052 and CS31082-001 agree very well with the r-process ratios predicted by \citet{Prantzos2020};
	\item the [Ag/Eu] ratios are close to the r-process one in the [Fe/H] $\lesssim -1.6$ stars, except HD~122563, HD~2796, and BD~+05 3098,
	 while they are close to the solar value in the $-1.6 \lesssim$ [Fe/H] $\le -0.56$ stars.  
\end{itemize}

Based on the r-residuals of \citet{Prantzos2020}, we conclude that production of silver was dominated by the main r-process throughout the metallicity range we investigated. For [Fe/H] $\gtrsim -1.6$, silver was also produced by the main s-process, however, its contribution cannot exceed the solar value 20.7\%.
With the exception for HD~122563, HD~2796, and BD~+05 3098, the stellar elemental ratios involving Ag do not suggest contributions from any additional nucleosynthetic processes. 
NLTE is essential for obtaining accurate observational constraints to future models of the Galactic Ag evolution.

\begin{acknowledgments}
The work was carried out under the state assignment of the Institute of astronomy of RAS and with the support of an internal grant of Herzen University (project No. 46-VG).
We made use of the MARCS, NIST, VALD, and ADS\footnote{http://adsabs.harvard.edu/abstract\_service.html} databases.
\end{acknowledgments}

\begin{contribution}
L.~M. and T.~S. have built the model atom for \ion{Ag}{1}, determined the solar and stellar Ag abundances, discussed the Galactic elemental ratios. S.~A.~Y. and V.~A.~V. have calculated the \ion{Ag}{1} + \ion{H}{1} collisions.
\end{contribution}

\facilities{VLT(UVES, HARPS)}

\software{\textsc{detail} \citep{Giddings81,Butler84,mash_fe}, \textsc{siu} \citep{Reetz}, \textsc{synthV\_NLTE} \citep{2019ASPC}, \textsc{BinMag} \citep{2018ascl.soft05015K}
}

\bibliography{atomic_data,mashonkina,nlte,pd,references}{}
\bibliographystyle{aasjournalv7}

\end{document}